\documentclass[11pt]{article}

\usepackage{graphicx} 
\usepackage{amsmath}
\usepackage{amsfonts}
\usepackage{amssymb}

\begin{document}

\title{Black Holes in the Ghost Condensate}
\author{Shinji Mukohyama\\
\\
{\it Department of Physics}\\
{\it and}\\
{\it Research Center for the Early Universe}\\
{\it The University of Tokyo, Tokyo 113-0033, Japan}\\
\\
{\it Jefferson Laboratory of Physics}\\
{\it Harvard University, Cambridge, Massachusetts 02138, USA}
}
\date{\today}

\maketitle

\abstract{%
We investigate how the ghost condensate reacts to black holes immersed
in it. A ghost condensate defines a hypersurface-orthogonal congruence
of timelike curves, each of which has the tangent vector
$u^{\mu}=-g^{\mu\nu}\partial_{\nu}\phi$. It is argued that the ghost
condensate in this picture approximately corresponds to a congruence of
geodesics. In other words, the ghost condensate accretes into a black
hole just like a pressure-less dust. Correspondingly, if the energy
density of the ghost condensate at large distance is set to an extremely
small value by cosmic expansion then the late-time accretion rate of the
ghost condensate should be negligible. The accretion rate remains very
small even if effects of higher derivative terms are taken into account, 
provided that the black hole is sufficiently large. It is also discussed
how to reconcile the black hole accretion with the possibility that the
ghost condensate might behave like dark matter. 
\begin{flushright}
  UTAP-512, RESCEU-2/05, HUTP-05/A0009
\end{flushright}
}

\section{Introduction and Summary}
\label{sec:intro}

Gravity at long distances shows us many interesting and mysterious
phenomena: flattening galaxy rotation curves, dimming supernovae, and so
on. These phenomena have been a strong motivation for the paradigm of
dark matter and dark energy, i.e. unknown components of the universe 
which show up only gravitationally. As we essentially do not know what
the dark matter and the dark energy are, however, it seems a healthy
attitude to consider the possibility that gravity at long distances
might be different from what we think we know.

This kind of consideration has been a motivation for attempts for IR
modification of gravity, e.g. massive gravity~\cite{Fierz-Pauli} and 
DGP brane model~\cite{DGP}. However, they are known to have a
macroscopic UV scale at around $1000$km, where effective field theories
break down~\cite{AGS,LPR}. This does not necessarily mean that these
theories cannot describe the real world, but implies that we need
non-trivial assumptions about the unknown UV completion. The recent
proposal of ghost condensation~\cite{paper1} evades at least this
problem and can be thought to be a step towards a consistent theory of
IR modification of general relativity.

In general, if we have scalar fields then there are many things we can 
play with them. In cosmology, inflation can be driven by the potential
part of a scalar field. It is also possible to drive inflation by the
kinetic part of a scalar field~\cite{k-inflation}. On the other hand,
scalar fields play important roles also in particle physics. A scalar
field is used for spontaneous symmetry breaking and to change force
laws in the Higgs mechanism. This is usually achieved by using a
potential whose global minimum is charged under the gauge symmetry. The
basic idea of ghost condensation is to break a symmetry and change a
force law by the kinetic part of a scalar field. In this sense the ghost
condensation can be considered as an analog of Higgs mechanism. Note
that modifying gravity force law via spontaneous symmetry breaking,
i.e. ghost condensation, is different from just adding a new matter in
the sense that the linearized gravity is modified even in Minkowski or
de Sitter background.

Since the ghost condensation modifies gravity, it is natural to ask the
question ``what happens to the ghost condensate when gravity is very
strong?'' Two such situations are in the early universe and near a black
hole. Effects of gravity in the early universe were already investigated
in refs.~\cite{paper1,paper2}. Hence, the next question would be ``what
happens in the other regime of strong gravity, namely near a black
hole?'' This is the subject of this paper. Other interesting topics
related to the ghost condensate include moving
sources~\cite{Dubovsky,Peloso-Sorbo}, nonlinear
dynamics~\cite{KRRZ,EJM,paper4},
cosmology~\cite{Senatore,Krause-Ng,Anisimov-Vikman}, galaxy rotation  
curve~\cite{Kiselev}, spin-dependent force~\cite{paper3}, and so on.

Before discussing the ghost condensate near a black hole, we begin with
briefly reviewing a well-known fact about observer-dependence of
gravitational force. A black-hole horizon forms when gravity is very
strong in the sense that even the degenerate pressure due to neutrons
cannot support the implosive gravitational force. However, as is
well-known, different observers feel different gravitational forces
since a force is defined by acceleration of an observer's
trajectory. This is particularly notable near a black-hole horizon. For
a static observer, the closer to the horizon the observer's position is,
the stronger the gravitational force is. Indeed, acceleration of a
static observer diverges at the horizon. On the other hand, for a freely
falling observer, a black-hole horizon is not a special point and
actually there is nothing divergent at the horizon. Indeed, the
acceleration of a freely-falling point-like observer vanishes by
definition. An extended object passing through a black hole horizon does
feel a tidal force due to the non-zero Riemann curvature, but the tidal
force is negligible for a sufficiently large black hole.

Now a ghost condensate in general defines a hypersurface-orthogonal
timelike vector field $u^{\mu}=-\partial^{\mu}\phi$. Thus, it is
possible to regard the ghost condensate as a hypersurface-orthogonal
congruence of timelike curves, each of which has the tangent vector
$u^{\mu}$. In this paper we shall argue that, when the ghost condensate
in this picture approximately corresponds to a congruence of geodesics,
the accretion rate of a ghost condensate into a black hole should be
negligible for a sufficiently large black hole. The essential reason for
the smallness of the accretion rate is the same as that for the
smallness of the tidal force acted on an extended object freely falling
into a large black hole.

In the rest of this paper, for simplicity we consider a scalar field
$\phi$ described by the action 
%============< EQUATION >==============%
%
\begin{equation}
 I = \int d^4x\sqrt{-g}
  \left[ P(X) - \frac{\alpha(\Box\phi)^2}{2M^2}\right],
  \label{eqn:action}
\end{equation}
%======================================%
where $X=-\partial^{\mu}\phi\partial_{\mu}\phi$ and the sign convention 
for the metric is $(-+\cdots +)$. Hereafter, we assume that $P'(M^4)=0$,
$P(M^4)=0$ and $P''(M^4)>0$, where the first equation just defines the
scale $M$, the second condition corresponds to zero cosmological
constant in the Higgs phase, and the third condition is required by the
absence of ghost in the Higgs phase. As shown in \cite{paper1}, $P'$ is
set to an extremely small value, or $X\to M^4$ by the expansion of the
universe.

In the following we investigate a ghost condensate interacting with a
black hole and present an approximate $P'=0$ solution, for which 
gravitational backreaction such as accretion rate is very
small. A.~Frolov~\cite{Frolov} considered different solutions with 
$P'\ne 0$, which correspond to congruences of non-geodesic (namely
accelerated) observers, and obtained a large accretion rate due to 
the stress-energy tensor of order $M^4P'$. However, as the accretion
proceeds, the energy density around the black hole, which is also of
order $M^4P'$, should decrease and $P'$ near the black hole should
approach to zero or a small value. Therefore, the accretion should slow
down because of shortage of energy and the steady-state accretion
claimed in ref.~\cite{Frolov} cannot be established. In the following we
shall obtain a much smaller accretion rate for an approximate $P'=0$
solution. Note that setting $P'\simeq 0$ is completely natural since the
expansion of the universe makes $P'$ extremely small.

We expect that the effects of the $\alpha$ term (corresponding somehow 
to the tidal force for a freely falling extended object) are small for a 
sufficiently large black hole. To be more precise, for a large enough 
black hole, there should be an approximate $X=M^4$ solution for which the
$\alpha$ term can be treated perturbatively. Thus, we first construct an
appropriate $X=M^4$ solution and then introduce deviation from it due to
the $\alpha$ term as a perturbation.

\begin{figure}
 \begin{center}
  \includegraphics[trim = 0 0 0 0 ,scale=0.4, clip]{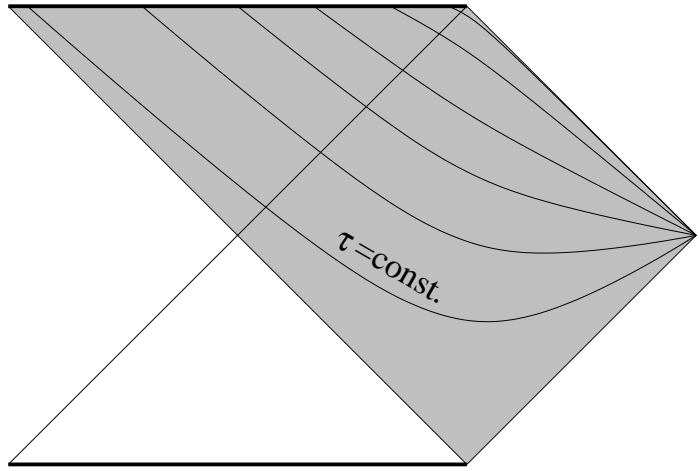}
 \end{center}
 \caption{\label{fig:tau-const}
 Constant-$\tau$ surfaces are drawn for the Gaussian normal coordinate
 system (\ref{eqn:metric-Gaussian}) of Schwarzschild metric. The
 coordinate system covers the shaded region. 
 }
\end{figure}

\begin{figure}
 \begin{center}
  \includegraphics[trim = 0 0 0 0 ,scale=0.4, clip]{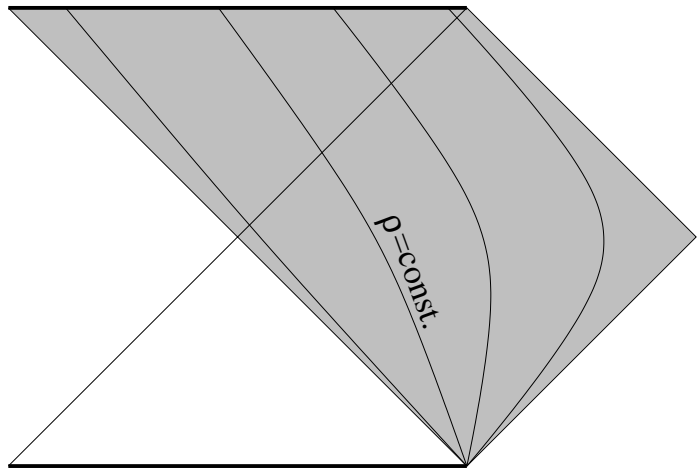}
 \end{center}
 \caption{\label{fig:rho-const}
 Constant-$\rho$ surfaces are drawn for the Gaussian normal coordinate
 system (\ref{eqn:metric-Gaussian}) of Schwarzschild metric. The
 coordinate system covers the shaded region. 
 }
\end{figure}

In a Gaussian normal coordinate system called Lemaitre reference
frame~\cite{Frolov-Novikov}, the Schwarzschild geometry with mass
parameter $m_0$ is written as
%============< EQUATION >==============%
%
\begin{equation}
 ds^2 = -d\tau^2 + \frac{d\rho^2}{a(\tau,\rho)}
  + \rho^2a^2(\tau,\rho)d\Omega^2, 
  \label{eqn:metric-Gaussian}
\end{equation}
%======================================%
where
%============< EQUATION >==============%
%
\begin{equation}
 a(\tau,\rho) = 
  \left[ 1
   -\frac{3\tau}{4m_0}\left(\frac{2m_0}{\rho}\right)^{3/2}\right]^{2/3}.
\end{equation}
%======================================%
This coordinate system was originally found by Lemaitre~\cite{Lemaitre}
and independently by Rylov~\cite{Rylov} and Novikov~\cite{Novikov}. For
completeness, the coordinate transformation from the standard coordinate
system to this one is given in Appendix~\ref{app:gaussian-coord}. In
particular the usual areal radius $r$ is given in this coordinate by
$r=\rho a$ so that the event horizon is located at $\rho
a=2m_0$. Manifestly, there is nothing bad on the future (black hole)
horizon and the coordinate system covers everywhere in the shaded region
in Figs.~\ref{fig:tau-const} and \ref{fig:rho-const}. The metric becomes
ill only on the curvature (physical) singularity at $\rho a=0$. Each
worldline with $\rho=const$ in Fig.~\ref{fig:rho-const} corresponds to
an observer freely falling into the black hole. With $\alpha=0$,
$\phi=M^2\tau$ satisfies the equation of motion and the Einstein
equation since $X=M^4$ implies that the stress energy tensor of $\phi$
vanishes. In this sense, this coordinate choice provides an analog of
the unitary gauge in flat spacetime. For this solution with $\alpha=0$,
$\Box\phi$ is regular outside the horizon $\rho a>r_g$: 
%============< EQUATION >==============%
%
\begin{equation}
 0 < \frac{\Box\phi}{M^3} = \frac{3}{2\rho a M}\sqrt{\frac{r_g}{\rho a}} 
  < \frac{3}{2Mr_g},
\end{equation}
%======================================%
where $r_g=2m_0$ is the Schwarzschild radius. Hence, the effect of the
term $-\alpha(\Box\phi)^2/2M^2$ in the action is suppressed by the small
factor
%============< EQUATION >==============%
%
\begin{equation}
 \epsilon = \frac{\alpha}{M^2r_g^2}. 
  \label{eqn:def-epsilon}
\end{equation}
%======================================%
In Sec.~\ref{sec:accretion} we shall perform perturbation with respect
to $\epsilon$, assuming that $\alpha=O(1)$ and that the black hole
radius $r_g$ is sufficiently larger than the microscopic length scale 
$\sqrt{\alpha}/M$. We shall see that the accretion rate is negligible as
expected. 
%============< FIGURE >==============%
%       tau-const.eps
%======================================%
%============< FIGURE >==============%
%       rho-const.eps
%======================================%

As pointed out in ref.~\cite{paper1}, deviation from $P'=0$ in
homogeneous, isotropic cosmology behaves exactly like dark
matter. Hence, it is also interesting to estimate the mass increase of a
black hole due to the accretion of ghost condensate with non-vanishing
$P'$, whose value at large distance is set by the energy density of dark
matter. Since the $\alpha$ term gives negligible contribution to the
accretion rate, we set $\alpha=0$ in this analysis.

Let us first estimate the maximum volume from which a black hole can
in principle swallow the energy density of the ghost condensate. 
For this purpose, suppose that a volume with radius $R$ at $t=0$
falls into a black hole and passes the black hole horizon with radius
$r_g$ at $t=T$. A crude estimate for the relation between $R$ and $T$ is
obtained by considering a fiducial collapsing FRW universe filled with
dark matter: $R\propto T^{2/3}$. The proportionality coefficient is 
guessed by dimensional analysis as 
%============< EQUATION >==============%
%
\begin{equation}
 \frac{R}{r_g} \sim \left(\frac{T}{r_g}\right)^{\frac{2}{3}}. 
\end{equation}
%======================================%
Thus, the maximum mass increase $\Delta M_{BH}$ during the time
interval $T$ is estimated as
%============< EQUATION >==============%
%
\begin{equation}
 \frac{\Delta M_{BH}}{M_{BH}} \sim 
  \frac{\rho_{\pi\infty}R^3}{M_{Pl}^2r_g}
  \sim \frac{\rho_{\pi\infty}}{\rho_{tot\infty}}\times (H_0T)^2,
  \label{eqn:mass-increase-nonzeroP'}
\end{equation}
%======================================%
where $\rho_{\pi\infty}$ ($\propto a^{-3}$) is the energy density of the
ghost condensate at large distance, $\rho_{tot\infty}\sim M_{Pl}^2H_0^2$
is the total energy density at large distance, and $H_0$ is the Hubble 
expansion rate today. A more systematic and detailed calculation is
given in Sec.~\ref{sec:nonzeroP'} and the result is qualitatively the
same, provided that the mass increase is for the Misner-Sharp energy on
the black hole horizon and that $T$ is replaced by the advanced time $v$
normalized at past null infinity. It is easy to see that the mass
increase $\Delta M_{BH}$ given by (\ref{eqn:mass-increase-nonzeroP'}) is
not too large. Indeed, even the integration over the age of the universe
($T\sim H_0^{-1}$) gives   
%============< EQUATION >==============%
%
\begin{equation}
 \left.\frac{\Delta M_{BH}}{M_{BH}}\right|_{T\sim H_0^{-1}}
  \sim \frac{\rho_{\pi\infty}}{\rho_{tot\infty}} < 1. 
\end{equation}
%======================================%

The rest of this paper is organized as follows. In
Sec.~\ref{sec:accretion} explains the main result of this paper, the
small accretion rate for an approximate $X=M^4$ solution. In
Sec.~\ref{sec:nonzeroP'} we estimate the mass increase of a black hole
in the case that the ghost condensate is dark matter. 
Sec.~\ref{sec:discussions} is devoted to discussions.

\section{Accretion rate}
\label{sec:accretion}

In this section we consider corrections to a Schwarzschild geometry due 
to the ghost condensate. In particular, we calculate corrections to the 
Misner-Sharp energy to estimate the mass increase of a black hole do to
accretion of the ghost condensate.

The variation of the action (\ref{eqn:action}) is
%============< EQUATION >==============%
%
\begin{equation}
 \delta I = \int d^4x\sqrt{-g}
  \left( \frac{1}{2}T^{\mu\nu}\delta g_{\mu\nu} + E_{\phi}\delta\phi
  \right),
\end{equation}
%======================================%
where
%============< EQUATION >==============%
%
\begin{eqnarray}
 T^{\mu\nu} & = & 2P'\partial^{\mu}\phi\partial^{\nu}\phi
  -\frac{\alpha\partial^{\mu}(\Box\phi)\partial^{\nu}\phi}{M^2}
  -\frac{\alpha\partial^{\mu}\phi\partial^{\nu}(\Box\phi)}{M^2}
  + \left[ P+\frac{\alpha(\Box\phi)^2}{2M^2}
     + \frac{\alpha\partial^{\rho}(\Box\phi)\partial_{\rho}\phi}{M^2}
    \right]g^{\mu\nu}, \nonumber\\
 E_{\phi} & = & 2\nabla^{\mu}(P'\nabla_{\mu}\phi)
  - \frac{\alpha\Box^2\phi}{M^2},
\end{eqnarray}
%======================================%
so that the relevant equations of motion is 
%============< EQUATION >==============%
%
\begin{equation}
 M_{Pl}^2G_{\mu\nu} = T_{\mu\nu}, \quad E_{\phi} = 0. 
\end{equation}
%======================================%

We expect that the effects of the $\alpha$ term are small for a 
sufficiently large black hole and is of order $O(\epsilon)$, where
$\epsilon$ is defined by (\ref{eqn:def-epsilon}). Hence, we first seek a
solution for $\alpha=0$ and introduce a non-zero $\alpha$ as a
perturbation later. In order to find a solution with $\alpha=0$, we
consider Schwarzschild spacetime, 
%============< EQUATION >==============%
%
\begin{eqnarray}
 g_{\mu\nu}dx^{\mu}dx^{\nu} & = & -f_0(r)dt^2 + \frac{dr^2}{f_0(r)} + 
  r^2d\Omega^2, \nonumber\\
 f_0(r) & = & 1-\frac{2m_0}{r},
  \label{eqn:Schwarzshild}
\end{eqnarray}
%======================================%
and seek a solution to the equation $X=M^4$ in this background. This is 
because, with $\alpha=0$, any configurations with $X=M^4$ satisfy not
only the $\phi$-equation of motion but also the Einstein equation. After
finding the solution with $\alpha=0$, we shall calculate the order
$O(\epsilon^1)$ corrections and see that the corrections due to the
non-zero $\alpha$ are small enough and that the expansion with respect
to $\epsilon$ makes sense.

To give an explicit solution to $X=M^4$, let us consider the ansatz 
%============< EQUATION >==============%
%
\begin{equation}
 \phi = M^2 \left[ t + g(r)\right],
\end{equation}
%======================================%
for which
%============< EQUATION >==============%
%
\begin{equation}
 \frac{X}{M^4} = \frac{1}{f_0}-(\partial_rg)^2f_0.
\end{equation}
%======================================%
Hence, $X=M^4$ can easily be solved to give
%============< EQUATION >==============%
%
\begin{equation}
 \phi = \phi_{\pm} \equiv
  M^2\left\{ t \pm 2m_0
      \left[2\sqrt{\frac{r}{2m_0}}
       +\ln\left(\frac{\sqrt{r}-\sqrt{2m_0}}
	    {\sqrt{r}+\sqrt{2m_0}}\right)\right]\right\} .
 \label{eqn:phi-sch}
\end{equation}
%======================================%
For this solution, $\Box\phi$ is finite except at $r=0$:
%============< EQUATION >==============%
%
\begin{equation}
 \frac{\Box\phi_{\pm}}{M^2} = \pm\frac{3}{2r}\sqrt{\frac{2m_0}{r}}.
\end{equation}
%======================================%
Note that the $+$ sign in (\ref{eqn:phi-sch}) is appropriate for a black
hole formed by gravitational collapse since, for the $+$ sign,
$\phi/M^2\sim v\equiv t+r^*$ near the horizon, where $v$ is the advanced 
time and $r^*=r+2m_0\ln(r/2m_0-1)$. On the other hand, the $-$ sign is
appropriate for a white hole. Hereafter, we choose the $+$ sign since we
are interested in a black hole.

Now let us treat the $\alpha$ term as a perturbation. For this
purpose we consider the spherically symmetric, time dependent ansatz
%============< EQUATION >==============%
%
\begin{eqnarray}
 g_{\mu\nu}dx^{\mu}dx^{\nu} & = & -f(t,r)e^{-2\delta(t,r)}dt^2
  + \frac{dr^2}{f(t,r)} + r^2d\Omega^2, \label{eqn:metric}\\
 \phi & = & \phi_+(t,r) + \pi(t,r), \label{eqn:phi}
\end{eqnarray}
%======================================%
where
%============< EQUATION >==============%
%
\begin{eqnarray}
 f(t,r) & = & 1-\frac{2m(t,r)}{r}, \nonumber\\
 m(t,r) & = & m_0 + m_1(t,r), \nonumber\\
 \delta(t,r) & = & 0 + \delta_1(t,r),
\end{eqnarray}
%======================================%
and consider $\pi$, $m_1$ and $\delta_1$ as quantities of order
$O(\epsilon)$, where $\epsilon$ is defined by (\ref{eqn:def-epsilon}).

In the order $O(\epsilon)$, the Einstein equation becomes
%============< EQUATION >==============%
%
\begin{eqnarray}
 \partial_r m_1 - 2m_0F'(r)\partial_tm_1 & = & 
  -\frac{9M^2}{8M_{Pl}^2}\left(1-\frac{m_0}{r}\right), \nonumber\\ 
 \partial_r\delta_1 & = & -\frac{1+2m_0/r}{r(1-2m_0/r)}\partial_rm_1
  -\frac{9M^2m_0(3-2m_0/r)}{8M_{Pl}^2r^2(1-2m_0/r)}, \nonumber\\
 \sqrt{\frac{2m_0}{r}}
  \left[\partial_r\pi - 2m_0F'(r)\partial_t\pi\right]
  & = & \frac{M^2\delta_1}{1-2m_0/r}
  -\frac{(1-2m_0/r)M_{Pl}^2}{2P_0''M^6r^2}\partial_r m_1
  + \frac{M^2(1+2m_0/r)}{r(1-2m_0/r)^2}m_1 \nonumber\\
 & &  + \frac{9m_0(1-2m_0/r)}{16M^4P_0''r^3},
\end{eqnarray}
%======================================%
where $P_0''=P''(M^4)$ and 
%============< EQUATION >==============%
%
\begin{equation}
 F(r) =
  \sqrt{\frac{r}{2m_0}}\frac{r+6m_0}{3m_0}
  + \ln\left(\frac{\sqrt{r}-\sqrt{2m_0}}{\sqrt{r}+\sqrt{2m_0}}\right),
  \quad
  F'(r) = \frac{1}{2m_0}\sqrt{\frac{r}{2m_0}}\frac{1}{1-2m_0/r}.
  \label{eqn:def-F}
\end{equation}
%======================================%
The solution to the first equation is 
%============< EQUATION >==============%
%
\begin{equation}
 \frac{m_1}{m_0} = \frac{9M^2}{4M_{Pl}^2}
  \left[-\frac{r}{2m_0}+\frac{1}{2}\ln\left(\frac{r}{2m_0}\right)
   + C(x_+)\right],\label{eqn:sol-m1-k4}
\end{equation}
%======================================%
where 
%============< EQUATION >==============%
%
\begin{equation}
 x_+ = F(r) + \frac{t}{2m_0},\label{eqn:def-xpm}
\end{equation}
%======================================%
and $C(x_+)$ is an arbitrary function of $x_+$. Note that the
$x_+=const.$ hypersurface is timelike and that
$x_+\sim v/2m_0+(5/3-2\ln 2)$ near the horizon, where $v=t+r_*$ is the
advanced time normalized at past null infinity, and
$r_*=r+2m_0\ln(r/2m_0-1)$. The finiteness of $m_1$ in the limit
$r\to\infty$ with initial, finite $t$ requires that the leading
asymptotic behavior of $C(x_+)$ for large positive $x_+$ should be 
%============< EQUATION >==============%
%
\begin{equation}
 C(x_+) \sim \left(\frac{3}{2}x_+\right)^{2/3}. 
  \label{eqn:asympto-C}
\end{equation}
%======================================%
This implies that the leading asymptotic behavior of $m_1$ on the black
hole horizon for large positive $v$ is
%============< EQUATION >==============%
%
\begin{equation}
 \frac{m_1}{m_0} \sim \frac{9M^2}{4M_{Pl}^2}
  \left(\frac{3v}{4m_0}\right)^{2/3}.
  \label{eqn:accretion-rate}
\end{equation}
%======================================%
This formula shows that the accretion rate $\partial_vm_1$ is very
small at late time. Indeed, the accretion rate is suppressed by the
factor $M^2/M_{Pl}^2$, reflecting the fact that there is no
gravitational backreaction in the decoupling limit $M_{Pl}^2\to\infty$.

For $M\sim 10MeV$, the Hubble expansion rate today $H_0$ is rewritten as
$M^3/M_{Pl}^2$. Thus, the formula (\ref{eqn:accretion-rate}) at
$v=H_0^{-1}$ is
%============< EQUATION >==============%
%
\begin{equation}
 \left.\frac{m_1}{m_0}\right|_{v=H_0^{-1}} \sim 
 \left(\frac{M_{Pl}}{M_{BH}}\right)^{2/3} \ll 1,
\end{equation}
%======================================%
where the black hole mass $M_{BH}$ is assumed to be much larger than the
Planck mass $M_{Pl}$ and we have set $\alpha=O(1)$. This result says
that the accretion of ghost condensate is negligible even if it is
integrated over the age of the universe.

\section{Accretion with $P'\ne 0$}
\label{sec:nonzeroP'}

We have obtained a negligible accretion rate,
assuming that $P'=0$ in the lowest order in the
$\epsilon$-expansion. This assumption is natural since the expansion of
the universe makes $P'$ extremely small, $P'\propto a^{-3}\to 0$
($a\to\infty$).

On the other hand, it is also interesting to consider non-zero $P'$ by
its own since the energy density associated with homogeneous,
non-vanishing $P'$ behaves exactly like dark matter. The linear
perturbation on top of the homogeneous background also behaves like dark
matter, but its nonlinear behavior remains to be seen~\cite{paper4}. In
this section we analyze how non-zero $P'$ changes the accretion of the
ghost condensate to a black hole.

Technically speaking, what we shall do in this section is the analysis
of spherically symmetric, time-dependent perturbation of the
Schwarzschild solution (\ref{eqn:Schwarzshild}) with
(\ref{eqn:phi-sch}). Throughout this section we set $\alpha=0$ since we
have already seen in the previous section that the accretion due
to non-zero $\alpha$ is negligibly small. We consider the spherically
symmetric, time dependent ansatz (\ref{eqn:metric}) with (\ref{eqn:phi})
and consider $\pi$, $m_1$ and $\delta_1$ as first-order quantities.

The perturbed Einstein equation is
%============< EQUATION >==============%
%
\begin{eqnarray}
 \partial_r m_1 - 2m_0F'(r)\partial_tm_1 & = & 0, \nonumber\\ 
 \partial_r\delta_1 & = & -\frac{1+2m_0/r}{r(1-2m_0/r)}\partial_rm_1, 
  \nonumber\\
 \sqrt{\frac{2m_0}{r}}
  \left[\partial_r\pi - 2m_0F'(r)\partial_t\pi\right]
  & = &  \frac{M^2\delta_1}{1-2m_0/r} 
  + \frac{M^2(1+2m_0/r)}{r(1-2m_0/r)^2}m_1 \nonumber\\
 & & -\frac{(1-2m_0/r)M_{Pl}^2}{2P_0''M^6r^2}\partial_rm_1,
  \label{eqn:Einstein-eq-without-alpha}
\end{eqnarray}
%======================================%
where $F(r)$ is defined by (\ref{eqn:def-F}). The solution to the first
equation is 
%============< EQUATION >==============%
%
\begin{equation}
 m_1 = \tilde{m}_1(x_+),\label{eqn:def-mtilde1}
\end{equation}
%======================================%
where $\tilde{m}_1$ is an arbitrary function and $x_+$ is defined by
(\ref{eqn:def-xpm}).

What we would like to know is the asymptotic behavior of
$m_1=\tilde{m}_1(x_+)$ for large $v$ on the black hole horizon, where 
$v$ is the advanced time. For this purpose we just have to specify a 
boundary condition at large $r$ with initial (finite) $t$ since 
$x_+\sim v/2m_0$ on the horizon and $x_+\sim (2/3)\cdot (r/2m_0)^{3/2}$
for large $r$ with finite $t$. For this purpose we give a formula
relating the perturbation $X_1$ of $X$ around $M^4$ to $\partial_t m_1$:
%============< EQUATION >==============%
%
\begin{eqnarray}
 X_1 & = & 
  2M^2\left\{-\sqrt{\frac{2m_0}{r}}
       \left[\partial_r\pi-2m_0F'(r)\partial_t\pi\right]
       + \frac{M^2\delta_1}{1-2m_0/r} 
       + \frac{M^2(1+2m_0/r)}{1-2m_0/r}m_1
      \right\}, \nonumber\\
 & = & 
  \frac{(1-2m_0/r)M_{Pl}^2}{P_0''M^4r^2}\partial_rm_1, \nonumber\\
 & = & 
  \frac{M_{Pl}^2}{\sqrt{2m_0}P''_0M^4}\frac{\partial_tm_1}{r^{3/2}}.
\end{eqnarray}
%======================================%
We have used the last equation in (\ref{eqn:Einstein-eq-without-alpha})
to obtain the second line and used the first equation in 
(\ref{eqn:Einstein-eq-without-alpha}) to obtain the last line. This
formula can be rewritten as a relation between $\tilde{m}_1'(x_+)$ and
the energy density of $\pi$ excitation $\rho_{\pi}$:
%============< EQUATION >==============%
%
\begin{equation}
 \rho_{\pi} = 2M^4 P_0''X_1 = 
  \frac{2M_{Pl}^2}{\sqrt{2m_0}r^{3/2}}\partial_tm_1
  =  \frac{2M_{Pl}^2}{\sqrt{2m_0}r^{3/2}}\frac{\tilde{m}'_1(x_+)}{2m_0},
\end{equation}
%======================================%
or
%============< EQUATION >==============%
%
\begin{equation}
 \frac{\tilde{m}'_1(x_+)}{2m_0}
  =  \frac{\sqrt{2m_0}r^{3/2}}{2M_{Pl}^2}\rho_{\pi}.
  \label{eqn:m1dash-rhopi}
\end{equation}
%======================================%

Now let us estimate the r.h.s. of (\ref{eqn:m1dash-rhopi}) at large $r$
with initial (finite) $t$. This gives the asymptotic behavior of
$\tilde{m}_1'(x_+)$ for large $x_+$ as
%============< EQUATION >==============%
%
\begin{equation}
 \frac{\tilde{m}'_1(x_+)}{2m_0}
  =  \frac{3m_0^2x_+}{M_{Pl}^2}\rho_{\pi\infty} \quad
  \mbox{ for } x_+\gg 1,
\end{equation}
%======================================%
where $\rho_{\pi\infty}$ is the energy density of $\pi$ excitation at
large $r$ with initial (finite) $t$. The l.h.s. of this equation is
actually equal to $\partial_v m_1$ on the black hole horizon
because of (\ref{eqn:def-mtilde1}). Thus, 
%============< EQUATION >==============%
%
\begin{equation}
 \left.\frac{\partial_v m_1}{m_0}\right|_{r=2m_0}
  =  \frac{3\rho_{\pi\infty}v}{2M_{Pl}^2}\quad
  \mbox{ for } v\gg 2m_0,
  \label{eqn:estimate-nonzeroP'}
\end{equation}
%======================================%
and integration w.r.t. $v$ gives
%============< EQUATION >==============%
%
\begin{equation}
 \left.\frac{m_1}{m_0}\right|_{r=2m_0}
  \simeq  \frac{3\rho_{\pi\infty}v^2}{4M_{Pl}^2}
  = \frac{9}{4}\frac{\rho_{\pi\infty}}{\rho_{tot\infty}}\times 
  (H_0v)^2 
  \quad \mbox{ for } v\gg 2m_0,
\end{equation}
%======================================%
where $\rho_{tot\infty}=3M_{Pl}^2H_0^2$ is the total energy density at
large distance and $H_0$ is the present Hubble expansion rate. This 
formula agrees with (\ref{eqn:mass-increase-nonzeroP'}) except for the 
$O(1)$-factor $9/4$, provided that the advanced time $v$ is replaced by
the fiducial cosmic time $T$.

\section{Discussions}
\label{sec:discussions}

A tachyon is considered to be sick in the context of particle mechanics,
but in field theory just indicates instability of a background. We
have considered a similar possibility called ghost
condensation~\cite{paper1} that a ghost field might be just an
indication of instability of a background and that it can condense to 
form a different background around which there is no ghost.

We have considered the question ``what happens to the ghost condensate
near a black hole?'' We have argued that the ghost condensate in this
picture approximately corresponds to a congruence of geodesics. In other
words, the ghost condensate accretes into a black hole just like a
pressure-less dust. Correspondingly, if the energy density of the ghost
condensate at large distance is set to an extremely small value by
cosmic expansion then the late-time accretion rate of the ghost
condensate should be negligible. The accretion rate remains very small
even if effects of higher derivative terms are taken into account,
provided that the black hole is sufficiently large. This has been
explicitly confirmed by a detailed calculation based on the perturbative
expansion with respect to a higher derivative term. The essential reason
for the smallness of the accretion rate due to the higher derivative
term is the same as that for the smallness of the tidal force acted on
an extended object freely falling into a large black hole. We have also
given an estimate of the mass increase of a black hole in the case that
the ghost condensate is dark matter and have shown that the accretion is
still slow.

In ref.~\cite{Frolov} A.~Frolov previously argued that the accretion
rate is huge, contrary to our result. One of the reasons for the
difference is that, while we have consistently taken into account
gravitational backreaction in the present paper, he neglected
gravitational backreaction. In ref.~\cite{Frolov}, by using solutions of
the equation of motion for the scalar field in a fixed geometry, a part
of the stress-energy tensor is calculated to give the accretion rate via
a part of the Einstein equation. However, this treatment neglects the
remaining components of the Einstein equation, which could completely
change both the geometry and the behavior of the scalar field. 
The large accretion rate is due to large $P'$, but $P'$ near the black
hole should decrease and approach to zero since the energy density,
which is of order $M^4P'$, decreases due to the accretion. Therefore,
the accretion should slow down because of shortage of energy and the
steady-state accretion claimed in ref.~\cite{Frolov} cannot be
established.

It is also interesting to notice that different scalar fields can behave
very differently near a black hole. We have found that the ghost
condensate near $P'=0$, i.e. within the validity of the low energy
effective field theory, accretes into a black hole just like a
pressure-less dust. On the other hand, Frolov and
Kofman~\cite{Frolov-Kofman} showed that a rolling (usual) scalar field
behaves like radiation near a black hole. It seems interesting to
classify the behaviors of different kinds of scalar fields near a black
hole and understand their behavior in more systematic way.

We have included the homogeneous component of the energy density (in
other words, cosmological energy density) of $\pi$ in the formula
(\ref{eqn:estimate-nonzeroP'}). For a black hole in a galaxy, one might
think that $\rho_{\pi\infty}$ should be replaced by some fraction of the
energy density of dark halo if the ghost condensate behaves as dark
matter in galaxies. In this case the mass increase in the unit of the
initial mass would become order unity within the galaxy dynamical
time. However, because of the following reason, we expect that accretion
should be slower. What makes the local density of $\pi$ in a galaxy
higher than the cosmological value should be nonlinear dynamics. In 
ref.~\cite{paper4} it is argued that caustics should form within the
Kepler time~\footnote{
See refs.~\cite{FKS,Felder-Kofman} for discussions about caustics for a
different system, a rolling tachyon.
} and that the ghost condensate should be described by a patchwork of
regular solutions. The size of each patchwork domain can be much smaller
than the size of the galaxy, depending on the dynamics. This means that
the ghost condensate averaged over galactic scales should have effective
rotation, i.e. angular momentum, around a black hole located in the
galaxy unless the initial condition is extremely fine-tuned. (Without
the patchwork, $\phi$ should be regular everywhere and there would be no
rotation: $\partial_{[\mu}\partial_{\nu]}\phi=0$.) The fine-tuning
required to make the effective rotation vanish is expected to be very
severe in the case of tiny ratio of the black hole size to the dark halo
size. With the effective rotation, it is not easy for the ghost
condensate to fall into a black hole straightforwardly. Thus, even if
the ghost condensate contributes to the dark halo significantly, we
expect that the dark halo component cannot accrete to a black hole
efficiently. On the other hand, the cosmological energy density of the
excitation of the ghost condensate can smoothly fall into a black hole
because of the absence of rotation. Therefore, the accretion rate for a
black hole in a galaxy should be between the one given in 
(\ref{eqn:estimate-nonzeroP'}) and the one which would be obtained by
setting $\rho_{\pi\infty}$ to the energy density of dark matter in
the galaxy. In other words, for a black hole in a galaxy, the mass
increase in the unit of the initial mass should become order unity in a
time scale between the galaxy dynamical time and the age of the
universe. It is worthwhile analyzing this issue in more detail.

\section*{Acknowledgements}

The author would like to thank N.~Arkani-Hamed, H.-C.~Cheng, M.~Luty,
J.~Thaler and T.~Wiseman for collaboration on ghost condensation and
their continuing encouragement. He is grateful to A.~Frolov, V.~Frolov, 
W.~Israel and L.~Kofman for many useful discussions and comments. 

\appendix

\section{Gaussian normal coordinate}
\label{app:gaussian-coord}

We can find a Gaussian normal coordinate system motivated by the
analysis in Sec.~\ref{sec:accretion}. First, let us consider $\phi_+$
and $x_+$ as time and space coordinates. We can calculate metric
components as  
%============< EQUATION >==============%
%
\begin{eqnarray}
 \partial^{\mu}\phi_+\partial_{\mu}\phi_+ & = & -M^4, \nonumber\\
 \partial^{\mu}x_+\partial_{\mu}x_+ & = & \frac{r}{(2m_0)^3},
  \nonumber\\ 
 \partial^{\mu}\phi_+\partial_{\mu}x_+ & = & 0.
\end{eqnarray}
%======================================%
Thus, the Schwarzschild metric is expressed as
%============< EQUATION >==============%
%
\begin{equation}
 ds^2  = -\frac{d\phi_+^2}{M^4} + \frac{(2m_0)^3}{r(\phi_+,x_+)}dx_+^2
  + r^2(\phi_+,x_+)d\Omega^2,
\end{equation}
%======================================%
where
%============< EQUATION >==============%
%
\begin{equation}
 r(\phi_+,x_+) \equiv 
  2m_0\left[\frac{3}{2}
       \left(x_+-\frac{\phi_+}{2m_0M^2}\right)\right]^{2/3}. 
\end{equation}
%======================================%
This coordinate system is nice in the sense that it covers everywhere in
the region $v>-\infty$ (namely, the relevant half of the Kruskal
extension) including the inside of the future (black hole)
horizon. However, it is not manifest how to deform this metric
continuously to the flat metric.

Hence, let us do one more coordinate transformation
$(\phi_+,x_+)\to (\tau,\rho)$, where
%============< EQUATION >==============%
%
\begin{equation}
 \tau \equiv \frac{\phi_+}{M^2}, \quad
  \rho \equiv 2m_0\left(\frac{3}{2}x_+\right)^{2/3}, \quad
  \tau < \tau_{max}(\rho) \equiv
  \frac{4m_0}{3}\left(\frac{\rho}{2m_0}\right)^{3/2}. 
\end{equation}
%======================================%
In this new coordinate system, the Schwarzschild solution is
%============< EQUATION >==============%
%
\begin{eqnarray}
 ds^2 & = & -d\tau^2 + \frac{d\rho^2}{a(\tau,\rho)}
  + \rho^2a^2(\tau,\rho)d\Omega^2, \nonumber\\
 \phi & = & M^2\tau,
\end{eqnarray}
%======================================%
where
%============< EQUATION >==============%
%
\begin{equation}
 a(\tau,\rho) = 
  \left[1
   -\frac{3\tau}{4m_0}\left(\frac{2m_0}{\rho}\right)^{3/2}\right]^{2/3}.
\end{equation}
%======================================%
This coordinate choice is an analog of the unitary gauge in flat 
spacetime. Actually, the metric becomes the flat metric in the 
$m_0\to 0$ limit. The unbroken shift symmetry is
%============< EQUATION >==============%
%
\begin{equation}
 \phi \to \phi + M^2\tau_0, \quad
 \tau \to \tau + \tau_0, \quad
  \left(\frac{\rho}{2m_0}\right)^{3/2} \to
  \left(\frac{\rho}{2m_0}\right)^{3/2} + \frac{3\tau_0}{4m_0}.
\end{equation}
%======================================%

There is nothing bad on the future (black hole)
horizon and the coordinate system covers everywhere in the region
$v>-\infty$ (the shaded region in Figs.~\ref{fig:tau-const} and
\ref{fig:rho-const}). The metric becomes ill only on the curvature
(physical) singularity at $\rho a=0$. As a consistency check it is easy
to calculate the Ricci tensor $R_{\mu\nu}$ and the Misner-Sharp energy
$M_{MS}$ for this metric as
%============< EQUATION >==============%
%
\begin{equation}
 R_{\mu\nu} = 0, \quad
 M_{MS} = \frac{\rho a}{2}
  \left[ 1 - \partial^{\mu}(\rho a)\partial_{\mu}(\rho a)\right]
  = m_0.
\end{equation}
%======================================%

\end{document}